\documentclass[%
reprint,
amsmath,amssymb,
aps,
pra,
floatfix,
]{revtex4-2}
\pdfoutput=1
\usepackage[colorlinks,allcolors=blue]{hyperref}
\usepackage{float}
\usepackage{bm}
\usepackage{graphicx}
\usepackage{dcolumn}
\usepackage[displaymath]{lineno}
\makeatletter
\let\LN@align\align
\let\LN@endalign\endalign
\renewcommand{\align}{\linenomath\LN@align}
\renewcommand{\endalign}{\LN@endalign\endlinenomath}
\let\LN@gather\gather
\let\LN@endgather\endgather
\renewcommand{\gather}{\linenomath\LN@gather}
\renewcommand{\endgather}{\LN@endgather\endlinenomath}

\makeatother

\begin{document}
\title{Nonreciprocity and mode conversion in a spatiotemporally modulated elastic wave circulator}
\author{Benjamin M. Goldsberry}
\affiliation{Applied Research Laboratories, The University of Texas at Austin, Austin, Texas 78758, USA}
\author{Samuel P. Wallen}
\author{Michael R. Haberman}
\affiliation{Applied Research Laboratories, The University of Texas at Austin, Austin, Texas 78758, USA}
\affiliation{Walker Department of Mechanical Engineering, The University of Texas at Austin, Austin, Texas 78712-1591, USA}
\date{\today}

\begin{abstract}
Acoustic and elastic metamaterials with time- and space-dependent material properties have received great attention recently as a means to break reciprocity for propagating mechanical waves, achieving greater directional control. One nonreciprocal device that has been demonstrated in the fields of acoustics and electromagnetism is the circulator, which achieves unirotational transmission through a network of ports. This work investigates an elastic wave circulator composed of a thin elastic ring with three semi-infinite elastic waveguides attached, creating a three-port network. Nonreciprocity is achieved for both longitudinal and transverse waves by modulating the elastic modulus of the ring in a rotating fashion. Two numerical models are derived and implemented to compute the elastic wave field in the circulator. The first is an approximate model based on coupled mode theory, which makes use of the known mode shapes of the unmodulated system. The second model is a finite element approach that includes a Fourier expansion in the harmonics of the modulation frequency and radiation boundary conditions at the ports. The coupled mode model shows excellent agreement with the finite element model and the conditions on the modulation parameters that enable a high degree of nonreciprocity are presented. This work demonstrates that it is possible to create an elastic wave circulator that enables nonreciprocal mode-splitting of an incident longitudinal wave into outgoing longitudinal and transverse waves.
\end{abstract}

\maketitle
\section{Introduction}

Acoustic and elastic wave reciprocity is a fundamental principle that requires the symmetry of propagating waves between two points in space when propagation takes place in linear time-invariant (LTI) media \cite{achenbach2003reciprocity,strutt1871}.
The simplest expression of reciprocity is that the propagation of waves from source to receiver will be identical in both phase and magnitude if the source and receiver are interchanged.
Current research in nonreciprocal wave propagation therefore aims to break this symmetry between source and receiver, enabling direction-dependent wave phenomena such as one-way wave propagation and vibration isolation \cite{nassar2020nonreciprocity}.
Various techniques that have been investigated in enabling nonreciprocity have often relied on violating the assumption of a LTI system.
For example, nonlinear wave propagation in geometrically-asymmetric media have been shown to exhibit nonreciprocity \cite{liang2009,mojahed2019tunable, moore2018nonreciprocity}.
In addition, spatiotemporal modulation of the material properties \cite{trainiti2016non, nassar2017modulated, sugino2020nonreciprocal,shen2019nonreciprocal,riva2020non, oudich2019space,goldsberry2019non,wallen2019} via the utilization of active elements \cite{popa2014NL_NR,zhai2019,chen2020active,shen2019nonreciprocal}, including modulation of paired loss-gain media in non-Hermitian systems \cite{koutserimpas2018nonreciprocal,rosa2020dynamics}, enables nonreciprocal effects for linear waves.

Many nonreciprocal mechanical systems have been inspired by analogous multi-port electronic devices \cite{caloz2018electromagnetic}. An isolator is a two-port system that allows for the transmission of waves in one direction, and has been realized for acoustic waves in systems that utilize fluid flow \cite{wiederhold2019nonreciprocal}, spatiotemporal modulation of material properties \cite{trainiti2016non}, and nonlinearity \cite{darabi2019broadband,liang2009}. 
The gyrator is another two-port system that manipulates the phase of a wave in one direction, and has recently been achieved in a waveguide with fluid flow \cite{zangeneh2018doppler}. Finally, circulators are three-port devices that transmit waves in a unirotational fashion, and have been realized using both fluid flow \cite{fleury2014sound} and spatiotemporal modulation of the volumes of coupled cavities \cite{fleury2015subwavelength}.
Nonreciprocal circulators, which are typically composed of three ports with three-fold rotational symmetry connected to coupled acoustic or elastic resonators, are the focus of this work. These devices are primarily utilized to isolate source and receiver channels for full-duplex communication devices \cite{fleury2015NR}. 
Previous works have limited the analysis of mechanical circulators to the acoustic (scalar) case \cite{fleury2014sound,fleury2015subwavelength}, lumped-element mechanical systems that only admit one unique resonance \cite{fleury2015subwavelength,vila2017bloch}, and elastic devices utilizing an angular momentum bias imposed by steady-state rotation \cite{beli2018mechanical}.
However, little attention has been given to more general elastic wave circulators with spatiotemporally modulated material properties, whose dynamics may be greatly enriched by coupling between modes, including multiple wave polarizations in the circulator and all ports.

We investigate an elastic wave circulator composed of a thin elastic ring connecting three semi-infinite elastic waveguides, creating a geometrically-symmetric three-port network.
Nonreciprocity is achieved for both longitudinal and transverse waves by modulating the elastic modulus of the ring in a rotating fashion.
Details of the circulator geometry and modulation of the elastic modulus is given in Sec.~\ref{sec:Geometry}.
Two numerical models are derived and implemented to compute the elastic wave field in the circulator: The first model is an approximate coupled mode model (CMM), presented in Sec.~\ref{sec:CoupledMode}, which is based on previous research investigating nonreciprocal vibrations in finite Euler beams \cite{goldsberry2020nonreciprocal}. The CMM is a reduced-order model of the circulator and is therefore computationally efficient, making the CMM well-suited for efficient design iteration.
However, the assumptions employed for the CMM impose restrictions on the circulator geometry, limiting its ability to capture the full design space. The second model, presented in Sec.~\ref{sec:FEM}, addresses these limitations by adapting a finite element approach (FEA) that was previously derived by the authors to explore nonreciprocal elastic wave propagation in modulated elastic metamaterials \cite{goldsberry2019non}. In contrast to the CMM, the FEA requires a discretized mesh of the circulator geometry and is therefore computationally expensive, which limits a thorough exploration of the circulator design space. We therefore calibrate the CMM with the FEA in Sec.~\ref{sec:results} in order to accurately utilize the CMM to investigate circulator designs, and we show with both models that the proposed elastic wave circulator exhibits a large degree of nonreciprocity and can act as a nonreciprocal Lamb wave mode converter. We then summarize the findings of this work in Sec.~\ref{sec:conclusion}.

\section{Model Definition} \label{sec:Geometry}
\begin{figure}
    \centering
    \includegraphics[width=1.0\columnwidth]{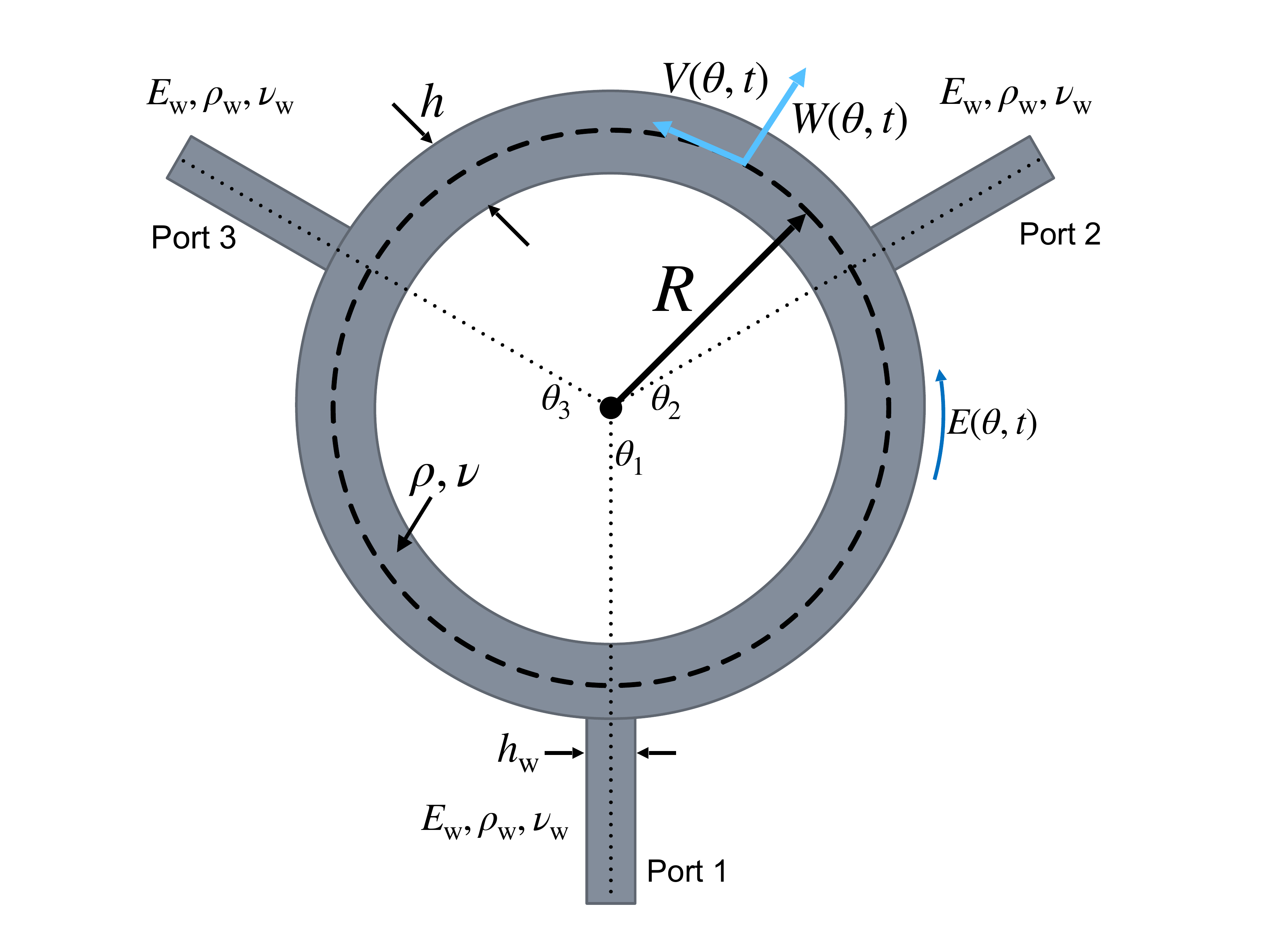}
    \caption{Elastic wave circulator geometry.}
    \label{fig:Geometry}
\end{figure}

Consider the elastic circulator configuration shown in Fig.~\ref{fig:Geometry}, which consists of an elastic ring attached to three semi-infinite, external elastic waveguides acting as input and output ports. The ring has constant radial thickness $h$ and constant radius of curvature $R$ at its mid-surface, such that the location of any point on the ring may be determined by the angle $\theta$. The semi-infinite waveguides each have thickness $h_\mathrm{w}$ and are attached at locations $\theta_1$, $\theta_2$, and $\theta_3$, separated by $120^\circ$, creating a three-port network with 3-fold axial symmetry.

The ring material is considered to be a linear, isotropic  solid with modulated Young's modulus $E(\theta,t)$, constant Poisson's ratio $\nu$, and constant mass density $\rho$. The corresponding material properties of the port waveguides are constants denoted by $E_\mathrm{w}$, $\nu_\mathrm{w}$, and $\rho_\mathrm{w}$, respectively, and may generally differ from the ring. The Young's modulus of the ring has the assumed form 
\begin{equation}
    E(\theta,t) = E_0 + E_\mathrm{m} E'(\theta,t), 
\end{equation}
where $E_0$ is the mean (static) Young's modulus, $E_\mathrm{m}$ is the modulation amplitude, and $E'(\theta,t)$ is the variation, or modulation, of the Young's modulus about the mean value.
The modulation is restricted to be a periodic function of time, $E'(\theta, t+T_\mathrm{m}) = E'(\theta, t)$, where $T_\mathrm{m}$ is the fundamental period of the modulation. The fundamental angular frequency of the modulation then follows as $\omega_\mathrm{m} = 2 \pi/T_\mathrm{m}$.

\section{Coupled mode model} \label{sec:CoupledMode}
\subsection{Equations of motion} \label{sec:CoupledModeEqns}

Here, we derive an approximate model for the circulator based on coupled mode theory. We begin by assuming sufficiently small drive frequencies for which the wavelength is much larger than the thickness $h$, such that the elastic ring can be well approximated as a curved Euler-Bernoulli beam with plane strain assumptions for the out-of-plane direction. 
Therefore, the complete in-plane displacement field of the ring can be described by the radial and tangential displacements, denoted by $W(\theta,t)$ and $V(\theta,t)$, respectively. The coupled partial differential equations of motion for the displacements are written in dimensionless form as \cite{graff2012wave}
\begin{multline}\label{eq:EOMW}
    \frac{R^2}{c_0^2} \frac{\partial^2  w}{\partial t^2} = \gamma^2 \frac{\partial^2}{\partial \theta^2}\left[ \bar E(\theta,t) \left(\frac{\partial v}{\partial \theta}
    - \frac{\partial^2 w}{\partial \theta^2}\right)\right] \\
    - \bar E(\theta,t) \left( w + \frac{\partial  v}{\partial \theta}\right) + f_r,
\end{multline}
\begin{multline}\label{eq:EOMV}
    \frac{R^2}{c_0^2} \frac{\partial^2 v}{\partial t^2} = \gamma^2\frac{\partial}{\partial \theta} \left[ \bar E(\theta,t) \left( \frac{\partial v}{\partial \theta} - \frac{\partial^2 w}{\partial \theta^2} \right) \right] \\
    +  \frac{\partial}{\partial \theta} \left[ \bar E(\theta,t) \left( \frac{\partial v}{\partial \theta} + w\right) \right] + f_\theta,
\end{multline}
where $w(\theta,t) = W(\theta,t)/U$ and $ v(\theta,t) = V(\theta,t)/U$ are the dimensionless radial and tangential displacements, respectively, and are normalized by an arbitrary reference displacement $U$. 
The relevant elastodynamic parameters in Eqns.~\eqref{eq:EOMW}-\eqref{eq:EOMV} are the dimensionless Young's modulus $\bar E(\theta,t) = E(x,t)/E_0$, the longitudinal sound speed $c_0 = \sqrt{E_0/\rho}$, and the dimensionless parameter 
\begin{equation}\label{eq:Gamma}
\gamma = R^{-1}\sqrt{D/(E_0 h)}, 
\end{equation}
which represents the flexural rigidity of the curved plate, where $D = E_0 h^3/[12(1-\nu^2)]$ is the flexural rigidity of a planar plate with the same thickness.
The external loading is defined by the forces $f_r$ and $f_\theta$, which represent normalized radial and tangential contributions, respectively, and include sources and radiation losses due to the attached elastic waveguide ports.
The external forces from the waveguide ports can be idealized as point forces if the external port waveguide thicknesses are much less than the ring thickness ( $h_\mathrm{w} \ll h$).
In this case, the external waveguide ports are assumed to be stress free in the tangential direction, i.e., $f_\theta = 0$.
Therefore, the only input wave considered in this work is a time-harmonic radial input force at port 1, which is caused by an incident longitudinal mode in the port waveguide.
The radiation losses due to the waveguide ports are approximated using the mechanical longitudinal impedance of a plate \cite{heckl1961compendium}, such that the normalized exterior forcing term is written as 
\begin{equation} \label{eq:ForcingR}
    f_r = \frac{2R F_{0}}{E_0 h U}\delta(\theta-\theta_1)e^{-i\omega_0 t} + \sum \limits_{q=1}^3 \frac{R \bar{Z}}{c_0} \frac{\partial \bar w}{\partial t} \delta(\theta-\theta_q),
\end{equation}
where the first term on the right-hand side of Eq.~\eqref{eq:ForcingR} is the force due to the incident longitudinal wave in port 1 ($\theta_1$ in Fig.~\ref{fig:Geometry}), and the second term on the right-hand side of Eq.~\eqref{eq:ForcingR} is the radiation loss from all three port waveguides ($\theta_1$, $\theta_2$, and $\theta_3$ in Fig.~\ref{fig:Geometry}).
Furthermore, $\delta(\theta)$ is the Dirac delta function, $\bar Z = \rho_\mathrm{w}c_\mathrm{w} h_\mathrm{w}/(\rho c_0 h)$ is the dimensionless radiation impedance, where  $c_\mathrm{w} = \sqrt{E_\mathrm{w}/[\rho_\mathrm{w} (1-\nu_\mathrm{w}^2)]}$ is the low-frequency limit of the longitudinal wave speed of the port waveguide, and $F_{0}$ is the amplitude of the time-harmonic input force with frequency $\omega_0$. 

\subsection{Coupled mode formulation}\label{sec:CoupledModeFormulation}
Due to the periodicity of $E'(\theta,t)$ in time, we can expand $E'(\theta,t)$ as a truncated Fourier series with $2P+1$ terms, such that the dimensionless Young's modulus is written as
\begin{equation} \label{eq:YoungsModFourierSeries}
    \bar E(\theta,t) = 1 + \alpha \sum \limits_{\substack{p=-P \\ p \neq 0}}^P \hat{E}^p(\theta) e^{-i p \omega_\mathrm{m}t},
\end{equation}
where $\hat{E}^p(\theta)$ are the Fourier components of $E'(\theta,t)$, and $\alpha = E_\mathrm{m}/E_0$ is the dimensionless modulation amplitude.
As a result of the time-periodic form of the Young's modulus and the incident wave, Eqns.~\eqref{eq:EOMW}-\eqref{eq:EOMV} possess steady-state solutions of the form
\begin{align}
\label{eq:SolnFormW}
    w(\theta,t) &= \sum \limits_{p=-P}^{P} \hat{w}^p(\theta)e^{-i(\omega_0 + p\omega_\mathrm{m})t}, \\
    \label{eq:SolnFormV}
    v(\theta,t) &= \sum \limits_{p=-P}^{P} \hat{v}^p(\theta) e^{-i(\omega_0 + p\omega_\mathrm{m})t},
\end{align}
where $\hat{w}^p(\theta)$ and $\hat{v}^p(\theta)$ are the angle-dependent radial and tangential displacements for each frequency component $(\omega_0 + p \omega_\mathrm{m})$.
Since the displacements of the ring must be continuous for all $\theta$, including the endpoints $w(0,t) = w(2\pi,t)$ and $v(0,t) = v(2\pi,t)$, the spatial profiles can be expanded as a truncated Fourier series with $2N+1$ terms and basis $\psi_n(\theta) = e^{i n \theta}$,
\begin{align}
    \hat{w}^p &= \sum \limits_{n=-N}^N \hat{w}^p_n \psi_n(\theta), \\
    \hat{v}^p &= \sum \limits_{n=-N}^N \hat{v}^p_n \psi_n(\theta),
\end{align}
which, when substituted in Eqns.~\eqref{eq:SolnFormW}-\eqref{eq:SolnFormV}, yields the total displacement solution
\begin{align} \label{eq:wTotalSoln}
    w(\theta,t) &= \sum \limits_{p=-P}^P \sum \limits_{n=-N}^N \hat{w}^p_n e^{i\left[n\theta - \left(\omega_0 + p \omega_\mathrm{m}\right)t\right]}, \\
    v(\theta,t) &= \sum \limits_{p=-P}^P \sum \limits_{n=-N}^N \hat{v}^p_n e^{i\left[n\theta - \left(\omega_0 + p \omega_\mathrm{m}\right)t\right]}. \label{eq:vTotalSoln}
\end{align}
Equations~\eqref{eq:wTotalSoln}-\eqref{eq:vTotalSoln} show that the total displacement solutions are a sum of clockwise ($n < 0$) and counter-clockwise ($n > 0$) propagating plane waves at the frequencies $(\omega_0 + p \omega_\mathrm{m})$ with amplitudes $\hat{w}^p_n$ and $\hat{v}^p_n$. 

Equations for the plane wave amplitudes are derived using a weak formulation of the equations of motion, which results in a symmetric system of equations for the plane wave amplitude values, and also allows for arbitrary and possibly discontinuous stiffness modulation profiles, such as the traveling modulation form considered in \cite{nassar2017modulated}. The weak formulation of the equations of motion is derived by multiplying Eqns.~\eqref{eq:EOMW}-\eqref{eq:EOMV} by $\psi^*_m(\theta) = e^{-i m \theta}$ and integrating over $\theta$:
\begin{multline} \label{eq:wInt}
    \frac{R^2}{c_0^2} \int \limits_{0}^{2\pi}\frac{\partial^2 w}{\partial t^2} \psi^*_m \, d\theta = \\
    \gamma^2 \int \limits_0^{2\pi} \frac{\partial^2}{\partial \theta^2}\left[ \bar E(\theta,t) \left(\frac{\partial v}{\partial \theta} - \frac{\partial^2 w}{\partial \theta^2}\right)\right] \psi^*_m\, d\theta \\
    - \int \limits_0^{2\pi} \bar E(\theta,t) \left(w + \frac{\partial v}{\partial \theta}\right) \psi^*_m\, d\theta + \int \limits_0^{2\pi}f_r \psi^*_m\, d\theta,
    \end{multline}
    \begin{multline}
    \frac{R^2}{c_0^2} \int \limits_0^{2\pi}\frac{\partial^2 v}{\partial t^2}\psi^*_m \, d\theta  = \\
    \gamma^2 \int \limits_0^{2\pi} \frac{\partial}{\partial \theta} \left[ \bar E(\theta,t) \left( \frac{\partial v}{\partial \theta} - \frac{\partial^2 w}{\partial \theta^2} \right) \right] \psi^*_m\, d\theta \\
    +  \int \limits_0^{2\pi}\frac{\partial}{\partial \theta} \left[ \bar E(\theta,t) \left( \frac{\partial v}{\partial \theta} + w\right) \right] \psi^*_m\, d\theta. \label{eq:vInt}
\end{multline}
Integration by parts is utilized twice on the first integral on the right hand side of Eq.~\eqref{eq:wInt}, and once on the first and second integrals on the right hand side of Eq.~\eqref{eq:vInt}, to yield
\begin{multline}\label{eq:wWF}
    \frac{R^2}{c_0^2} \int \limits_{0}^{2\pi}\frac{\partial^2 w}{\partial t^2} \psi^*_m \, d\theta  = \gamma^2 \int \limits_0^{2\pi} \bar E(\theta,t) \left(\frac{\partial v}{\partial \theta} - \frac{\partial^2 w}{\partial \theta^2}\right) \frac{\partial^2 \psi^*_m}{\partial \theta^2}\, d\theta \\
    - \int \limits_0^{2\pi} \bar E(\theta,t) \left(w + \frac{\partial v}{\partial \theta}\right) \psi^*_m\, d\theta + \int \limits_0^{2\pi}f_r \psi^*_m\, d\theta, 
    \end{multline}
    \begin{multline}
    \frac{R^2}{c_0^2} \int \limits_0^{2\pi}\frac{\partial^2 v}{\partial t^2}\psi^*_m \, d\theta = -\gamma^2 \int \limits_0^{2\pi} \bar E(\theta,t) \left( \frac{\partial v}{\partial \theta} - \frac{\partial^2 w}{\partial \theta^2} \right) \frac{\partial \psi^*_m}{\partial \theta}\, d\theta \\
    - \int \limits_0^{2\pi} \bar E(\theta,t) \left( \frac{\partial v}{\partial \theta} + w\right) \frac{\partial \psi^*_m}{\partial \theta}\, d\theta . \label{eq:vWF}
\end{multline}
Note that the boundary terms that arise from integration by parts vanish due to the spatial periodicity of the displacement solution and Young's modulus.
Finally, the equations for the plane wave amplitudes are found by substituting Eqns.~\eqref{eq:YoungsModFourierSeries}, \eqref{eq:wTotalSoln}, and \eqref{eq:vTotalSoln} into Eqns.~\eqref{eq:wWF}-\eqref{eq:vWF} and utilizing the orthogonality of the Fourier series in time and space. The resulting expressions can be written compactly, in matrix-vector form, as
\begin{widetext}
\begin{multline} \label{eq:HarmonicModalAmpEqns}
\left(\left(\bar{\omega}_0+p\bar{\omega}_\mathrm{m}\right)^2\begin{bmatrix}
\mathbf{I} & 0 \\
0 & \mathbf{I}
\end{bmatrix}
- \begin{bmatrix}
\mathbf{K}^{0}_\mathrm{W,W} & \mathbf{K}^{0}_\mathrm{W,V} \\
\mathbf{K}^{0}_\mathrm{V,W} & \mathbf{K}^{0}_\mathrm{V,V}
\end{bmatrix} +
i\bar{Z}\left(\bar{\omega}_0 + p\bar{\omega}_\mathrm{m}\right)\begin{bmatrix}
 \mathbf{C} & 0 \\
0 & 0
\end{bmatrix}
\right)
\begin{bmatrix}
\boldsymbol{w}^p\\
\boldsymbol{v}^p
\end{bmatrix} - \alpha \sum \limits_{\begin{smallmatrix} s = -P \\ s \neq p\end{smallmatrix}}^P \begin{bmatrix}
\mathbf{K}^{p-s}_\mathrm{W,W} & \mathbf{K}^{p-s}_\mathrm{W,V} \\
\mathbf{K}^{p-s}_\mathrm{V,W} & \mathbf{K}^{p-s}_\mathrm{V,V}
\end{bmatrix}
\begin{bmatrix}
\boldsymbol{w}^s \\
\boldsymbol{v}^s
\end{bmatrix}
 = \delta_{p0}\begin{bmatrix}
\boldsymbol{F} \\
0
\end{bmatrix}, \\
\quad
p \in [-P,P],
\end{multline}
\end{widetext}
where $\bm{w}^p = [\hat w^p_{-N}, \hat w^p_{-N+1}, ..., \hat w^p_N]^\mathrm{T}$ and $\bm{v}^p = [\hat v^p_{-N}, \hat v^p_{-N+1}, ..., \hat v^p_N]^\mathrm{T}$ are the vectors of plane wave amplitudes; $\bar \omega_0 = R\omega_0/c_0$ and $\bar \omega_\mathrm{m} = R \omega_\mathrm{m}/c_0$ are the dimensionless drive and modulation frequencies, respectively;
and the matrix $\mathbf{C}$, which represents the radiation loss in the ring due to the port waveguides, has the components
$
    C_{mn} = (2 \pi)^{-1}\sum \limits_{q=1}^3 e^{in\theta_q} e^{-i m \theta_q}.
$
The forcing vector $\boldsymbol{F}$ in Eq.~\eqref{eq:HarmonicModalAmpEqns} has the values $F_n = (2\pi)^{-1}e^{-i n \theta_\mathrm{s}}$, and the coupling matrices are
\begin{subequations}
\begin{multline} \label{eq:CouplingMatrixTerms}
    \left(K^{p-s}_{W,W}\right)_{mn} = -\gamma^2\int \limits_0^{2\pi} \hat{E}^{p-s}(\theta) \frac{\partial^2 \psi_n}{\partial \theta^2}\frac{\partial^2 \psi_m^*}{\partial \theta^2}\, d\theta \\
    - \int \limits_0^{2\pi} \hat{E}^{p-s}(\theta) \psi_n \psi_m^*\, d\theta,
    \end{multline}
    \begin{multline}
    \left(K^{p-s}_{V,W}\right)_{mn} = \left(K^{p-s}_{W,V}\right)_{mn} = \\
    \gamma^2\int \limits_0^{2\pi} \hat{E}^{p-s}(\theta) \frac{\partial \psi_n}{\partial \theta}\frac{\partial^2 \psi_m^*}{\partial \theta^2}\, d\theta - \int \limits_0^{2\pi} \hat{E}^{p-s}(\theta) \frac{\partial \psi_n}{\partial \theta} \psi_m^* \, d\theta,
    \end{multline}
    \begin{multline}
    \left(K^{p-s}_{V,V}\right)_{mn} = -(\gamma^2+1) \int \limits_0^{2\pi} \hat{E}^{p-s}(\theta) \frac{\partial \psi_n}{\partial \theta} \frac{\partial \psi_m^*}{\partial \theta}.
\end{multline}
\end{subequations}
The matrices inside the parentheses in Eq.~\eqref{eq:HarmonicModalAmpEqns} are the so-called mass, stiffness, and loss matrices, respectively, and are identical to the dynamic matrices for the unmodulated system.
The terms in Eq.~\eqref{eq:HarmonicModalAmpEqns} proportional to $\alpha$ arise due to the stiffness modulation, and represent the coupling between tangential and longitudinal wave modes within the ring at each frequency $(\bar \omega_0 + p \bar \omega_\mathrm{m})$.
Equation \eqref{eq:HarmonicModalAmpEqns} is numerically solved given the number of retained spatial plane waves $N$, frequencies $P$, and dimensionless parameters $(\bar \omega_0,\bar \omega_\mathrm{m}, \bar Z, \alpha)$. 
The results are shown in Sec.~\ref{sec:results}, and are compared with a finite element approach derived in Sec.~\ref{sec:FEM}.

\section{Finite element approach} \label{sec:FEM}
While the coupled mode model derived in Sec.~\ref{sec:CoupledMode} is computationally efficient, it has a number of assumptions, namely that the curvature $(h/R)$ and drive frequency must be sufficiently small such that Euler-Bernoulli theory holds. In addition, the port waveguide thickness must be much smaller than the ring thickness (\textit{i.e.}, $h_\mathrm{w}/h \ll 1$) such that the forcing and radiation impedance approximations in Eq.~\eqref{eq:ForcingR} are valid.
To alleviate these assumptions and obtain a model with applicability to more general geometries, we also develop a finite element approach (FEA), which numerically solves the continuum elastodynamic equations and takes into account the exact coupling between the ring and exterior waveguide ports. To this end, we extend the FEA previously derived in Ref. \cite{goldsberry2019non} to the present case. The primary modification that must be made is to enforce outgoing elastic waves for the reflected and transmitted fields that propagate in the waveguide ports, which is accomplished by adapting a mode-matching technique \cite{astley1996fe}.

\subsection{Derivation}\label{sec:FEM-Deriv}
The weak formulation for the elastodynamic equations with spatiotemporally-modulated elastic moduli is written in indicial notation, with the Einstein summation convention applied to subscript indices only, as \cite{goldsberry2019non}
\begin{multline} \label{eq:FEMWF}
    \sum \limits_{p,s=-P}^P - \int \limits_\Omega \hat{v}^p_{i,j}\hat{L}^{p-s}_{ijkl}\hat{u}_{k,l}\, d\Omega + \left(\omega_0 + p \omega_\mathrm{m}\right)^2 \int \limits_\Omega \rho \hat{u}^p_i \hat{v}^p_i\, d\Omega \\
    + \sum \limits_{q=1}^3\int \limits_{\Gamma_q} \hat{L}^0_{ijkl} \hat{u}^{p,\mathrm{port}}_{k,l} \hat{v}^p_i n_j\, dS = -\int \limits_{\Gamma_{\mathrm{port 1}}} \hat{L}^0_{ijkl} \hat{u}^{0,\mathrm{inc}}_{k,l} \hat{v}^0_i n_j\, dS,
\end{multline}
where $\hat{u}^p_i$ and $\hat{v}^p_i$ are the trial and test displacements, respectively, at frequency $(\omega_0 + p \omega_\mathrm{m})$, $\Omega$ represents the union of the ring and waveguide domains in Fig.~\ref{fig:Geometry}, and $\Gamma_q$, $q=1,2,3$ are the three exterior waveguide boundaries caused from the truncation of the semi-infinite waveguides to create a finite computational domain. The material properties in Eq.~\eqref{eq:FEMWF} include the mass density $\rho(\bm{x})$ and the fourth-order tangent modulus tensor 
\begin{equation}
    \hat{L}^s_{ijkl}(\bm{x}) = \hat{\lambda}^s(\bm{x}) \delta_{ij}\delta_{kl} + \hat{\mu}^s(\bm{x})\left(\delta_{ik}\delta_{jl} + \delta_{il}\delta_{jk}\right),
\end{equation}
where $\hat{\lambda}^s(\bm{x}) = \hat{E}^s(\bm{x})\nu/\left[(1+\nu)(1-2\nu)\right]$ and $\hat{\mu}^s(\bm{x}) = \hat{E}^s(\bm{x})/\left[2(1+\nu)\right]$ are the Fourier-transformed first and second Lam\'e parameters, respectively. We note that the $\hat{E}^s(\bm{x})$ functions are defined in Eq.~\eqref{eq:YoungsModFourierSeries} with $\theta = \mathrm{atan2}(y,x)$ and $\bm{x} = [x,y]^\mathrm{T}$. 
The unknown displacements at the port waveguide boundaries at each frequency, $\hat{\bm u}^{p,\mathrm{port}}$, must be determined such that the waves propagating away from the ring in the port waveguides remain purely outgoing and do not reflect from the computational boundary. 
Given the plane-strain assumptions of the model, we assume that the displacement field in the port waveguides can be completely represented as a linear combination of Rayleigh-Lamb waves \cite{graff2012wave}. 
The drive frequencies considered in this work are limited to be below the cut-on frequencies of the higher-order Lamb modes, such that only the zero order symmetric ($S_0$) and antisymmetric ($A_0$) Lamb waves propagate in the port waveguides.
The components of the displacement field for the symmetric Lamb wave, $\bm{\psi}^\mathrm{sym}$, are \cite{graff2012wave}
\begin{align}
    \psi^\mathrm{sym}_x &= i \left(\beta \cos(\beta y) + \frac{B}{C} \xi \cos(\eta y)\right) e^{i \xi x}, \\
    \psi^\mathrm{sym}_y &= \left(\xi \sin(\beta y) - \eta \frac{B}{C} \sin (\eta y)\right)e^{i \xi x}, \\
    \frac{B}{C} &= \frac{(\xi^2 - \beta^2) \sin(\beta h/2)}{2 \xi \eta \sin(\eta h/2)},
\end{align}
where $\xi$ is the wavenumber, 
 $\eta^2 = \omega^2/c_1^2 - \xi^2$, $\beta^2 = \omega^2/c_2^2 - \xi^2$, $c_1 = \sqrt{(\lambda + 2\mu )/\rho}$ is the bulk longitudinal wave speed of the waveguide material, and $c_2 = \sqrt{\mu/\rho}$ is the bulk transverse wave speed in the same material. 
Similarly, the components of the displacement field for the antisymmetric Lamb wave, $\bm{\psi}^\mathrm{asym}$, are
\begin{align}
    \psi^\mathrm{asym}_x &= -i \left(\beta \sin(\beta y) - \frac{A}{D} \xi \sin(\eta y)\right) e^{i \xi x}, \\
    \psi^\mathrm{asym}_y &= \left(\xi \cos(\beta y) + \eta \frac{A}{D} \cos (\eta y)\right)e^{i \xi x}, \\
    \frac{A}{D} &= -\frac{(\xi^2 - \beta^2)\cos(\beta h/2)}{2 \xi \eta \cos(\eta h/2)}.
\end{align}
The wavenumber $\xi$ satisfies the well-known Rayleigh-Lamb dispersion relation 
\begin{equation} \label{eq:RayleighLambDispersion}
    \frac{\tan(\beta h/2)}{\tan (\eta h/2)} + \left[\frac{4 \eta \beta \xi^2}{(\xi^2 - \beta^2)^2}\right]^{\pm 1} = 0,
\end{equation}
where the power of the second term on the left-hand side of Eq.~\eqref{eq:RayleighLambDispersion} is $+1$ for the symmetric Lamb wave and $-1$ for the antisymmetric Lamb wave.
The total displacement at each output port is therefore $\bm{u}^{p,\mathrm{port}} = A^{p,\mathrm{sym}} \bm{\psi}^{p,\mathrm{sym}} + A^{p,\mathrm{asym}}\bm{\psi}^{p,\mathrm{asym}}$, where $A^{p,\mathrm{sym}}$ and $A^{p,\mathrm{asym}}$ are the amplitude coefficients for the symmetric and antisymmetric Lamb waves, respectively, at the frequency $(\omega_0 + p\omega_\mathrm{m})$. 

The Lamb wave amplitudes are determined by creating an auxiliary equation to Eq.~\eqref{eq:FEMWF} that weakly enforces the continuity of displacement at each frequency. 
This is accomplished by forming a weighted inner product equation with operator $W$, $\left< W \bm{\psi}^{p}, \bm{u}^p - \bm{u}^{p, \mathrm{port}}\right> = 0$, where the weighting operator is chosen to be the integrand of the integrals over the port faces in Eq.~\eqref{eq:FEMWF} \cite{astley1996fe}. Expanding $\bm{u}^{p,\mathrm{port}}$ in terms of the symmetric and antisymmetric Lamb wave shapes yields the auxiliary equations
\begin{widetext}
\begin{align} \label{eq:AuxEqns1}
    \int \limits_{\Gamma_q}L^0_{ijkl} \psi_{k,l}^{p,\mathrm{sym}}n_j  \left(\hat{u}^p - A^{p,\mathrm{sym}} \psi^{p,\mathrm{sym}}\right)_{i}\, dS +  \int \limits_{\Gamma_q} L^0_{ijkl} \psi_{k,l}^{p,\mathrm{sym}}n_j \left(\hat{u}^p - A^{p,\mathrm{asym}} \psi^{p,\mathrm{asym}}\right)_{i}\, dS = 0, \\
     \int \limits_{\Gamma_q}L^0_{ijkl} \psi_{k,l}^{p,\mathrm{asym}}n_j \left(\hat{u}^p - A^{p,\mathrm{sym}}\psi^{p,\mathrm{sym}}\right)_{i}\, dS +  \int \limits_{\Gamma_q} L^0_{ijkl} \psi_{k,l}^{p,\mathrm{asym}}n_j \left(\hat{u}^p - A^{p,\mathrm{asym}}\psi^{p,\mathrm{asym}}\right)_{i}\, dS = 0, \label{eq:AuxEqns2}
\end{align}
\end{widetext}
for each waveguide boundary $\Gamma_q$.
Finally, an augmented system system of equations is solved for the finite element displacement field and the port Lamb wave amplitudes at each frequency,
\begin{equation} \label{eq:FEMAugSystemEqns}
    \begin{bmatrix}
    \omega_0^2 \mathbf{M} + \omega_0 \mathbf{C} + \mathbf{K} & \mathbf{\Gamma} \\
    \mathbf{\Gamma}^\mathrm{H} & \mathbf{\Lambda}
    \end{bmatrix}
    \begin{bmatrix}
    \bm{u} \\
    \bm{A}
    \end{bmatrix} = 
    \begin{bmatrix}
    \tilde{\bm{F}} \\
    \bm{S}
    \end{bmatrix},
\end{equation}
where $\bm{u} = [\bm{u}^{-P}, \bm{u}^{-P+1}, ..., \bm{u}^P]^\mathrm{T}$ is an augmented vector containing the finite element discretized displacements at each frequency, and $\bm{A} = [\bm{A}^{-P}, \bm{A}^{-P+1}, ..., \bm{A}^P]^\mathrm{T}$ is an augmented vector of the symmetric and antisymmetric Lamb wave amplitudes at each port waveguide and frequency. The matrices  $\mathbf{M}$, $\mathbf{C}$, and $\mathbf{K}$ in Eq.~\eqref{eq:FEMAugSystemEqns} are from the discretization of the integrals over $\Omega$ in Eq.~\eqref{eq:FEMWF} and are given in detail in \cite{goldsberry2019non}. The matrix $\mathbf \Gamma$ in Eq.~\eqref{eq:FEMAugSystemEqns} is from the integrals over the waveguide port surfaces on the left-hand side in Eq.~\eqref{eq:FEMWF}, where the superscript $\mathrm{H}$ denotes Hermitian transposition, and the matrix $\mathbf{\Lambda}$ is formed from the integrals proportional to the Lamb wave amplitude coefficients in Eqns.~\eqref{eq:AuxEqns1}-\eqref{eq:AuxEqns2}. Finally, the source terms on the right-hand side of Eq.~\eqref{eq:FEMAugSystemEqns} due to the incident wave are given by
\begin{align}
    \tilde{\bm{F}} &= \int \limits_{\Gamma_1} L^0_{ijkl} u^{\mathrm{inc}}_{k,l} v^0_i n_j\, dS, \\
    \bm{S} &= \int \limits_{\Gamma_1} L^0_{ijkl} u^{p,\mathrm{sym}}_{k,l}u_i^{\mathrm{inc}} n_j\, dS.
\end{align}
It is assumed in this work that the incident field is a symmetric Lamb wave, $\bm{u}^{\mathrm{inc}} = \bm{\psi}^\mathrm{sym}$, which is consistent with the input force for the coupled mode model discussed in Sec.~\ref{sec:CoupledModeEqns}.

\section{Results} \label{sec:results}
\begin{table}
\centering
    \begin{tabular}{|c|c|}
    \hline
        \textbf{Property} & \textbf{Value} \\ \hline
        $h/R$ & $0.05$ \\ \hline
        $h_\mathrm{w}/h$ & $0.2$ \\ \hline
        $E_\mathrm{0}$ & $195 \, \text{GPa}$ \\ \hline
        $E_\mathrm{w}$ & $12 \, \text{MPa}$ \\ \hline
        $\nu$ & $0.28$ \\ \hline
        $\nu_\mathrm{w}$ & $0.4$ \\ \hline
        $\rho$ & $7700 \, \text{kg}/\text{m}^3$\\ \hline
        $\rho_\mathrm{w}$ & $1000 \, \text{kg}/\text{m}^3$\\ \hline
    \end{tabular}
    \caption{Material and geometric properties for the CMM and FEA.}
    \label{table:1}
\end{table}
\begin{figure}
    \centering
    \includegraphics[width=1\columnwidth]{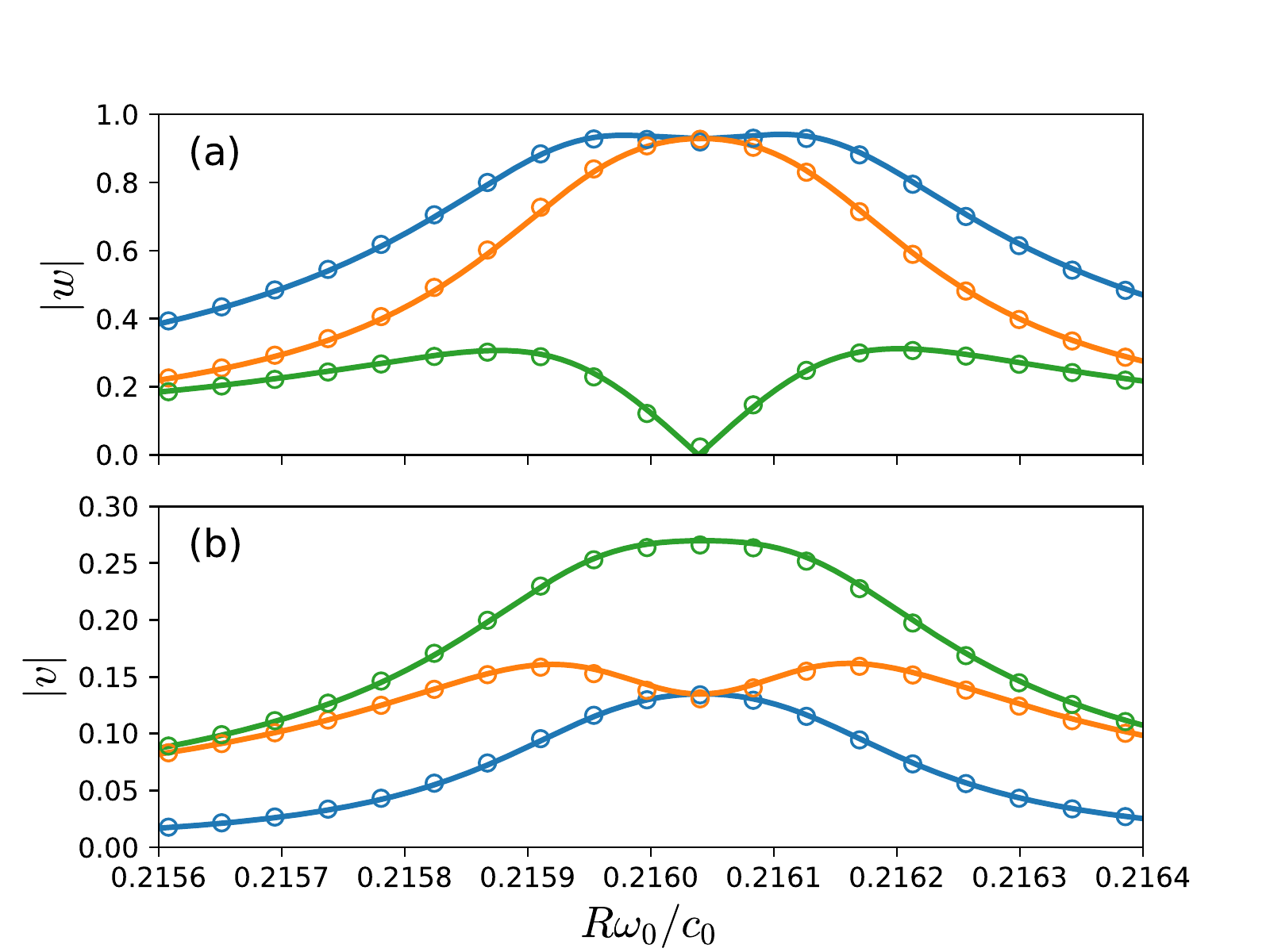}
    \caption{Magnitude of the (a) radial displacement, and (b) tangential displacement of the ring at the waveguide port locations for spatiotemporal parameters $(\alpha,\bar \omega_\mathrm{m}) = (0.2, 0.016)$. The blue, orange, and green curves denote the displacement magnitude at ports 1, 2, and 3, respectively. Furthermore, the solid lines and open circles are the results from the CMM and FEA, respectively.}
    \label{fig:FEM_CM_Comparison}
\end{figure}
\begin{figure*}
    \centering
    \includegraphics[width=0.8\textwidth]{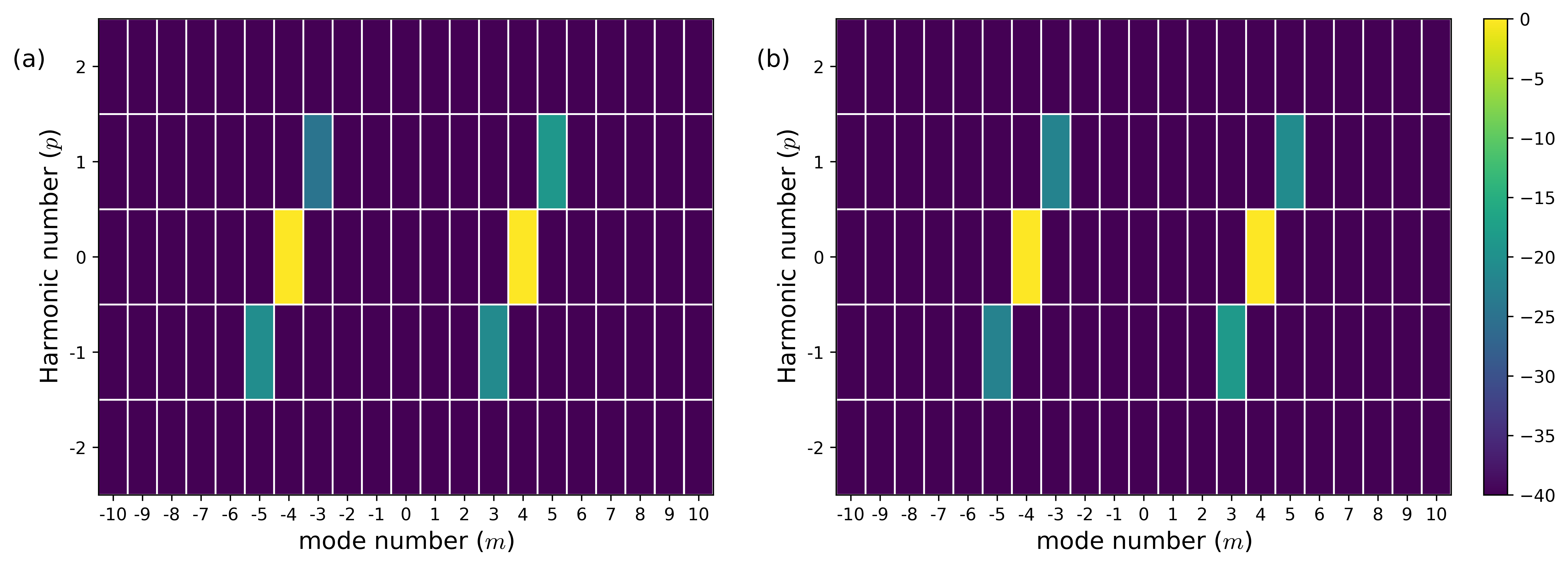}
    \caption{Absolute square of the harmonic modal amplitudes for (a) $|\hat{w}^p_m|^2$ and (b) $|\hat{v}^p_m|^2$ normalized by the absolute square of the maximum harmonic modal amplitude in decibels for spatiotemporal parameters $(\alpha,\bar \omega_\mathrm{m}) = (0.2, 0.016)$ and drive frequency $\bar \omega_0 = 0.21604$.}
    \label{fig:AmpMatrix}
\end{figure*}
\begin{figure*}
    \centering
    \includegraphics[width=1.5\columnwidth]{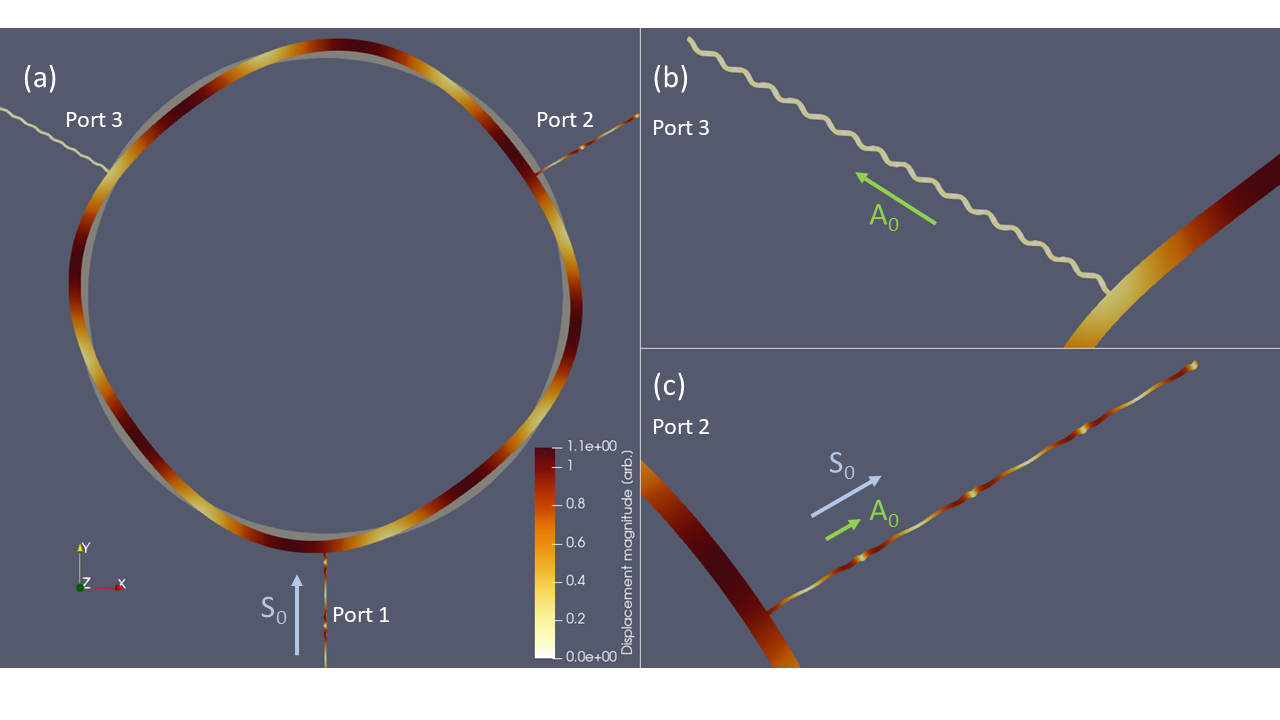}
    \caption{Finite element displacement solution for spatiotemporal parameters $(\alpha,\bar \omega_\mathrm{m}) = (0.2, 0.016)$ at the drive frequency $\bar{\omega}_0 = 0.21604$. The blue and green arrows denote the symmetric ($S_0$) and antisymmetric ($A_0$) Lamb wave, respectively, where the magnitude of the arrows qualitatively designate the relative contributions of each Lamb wave within the port waveguides. (a) Top-down view of the circulator. An incident symmetric lamb wave is incident to the circulator in port 1. (b) Zoom-in plot of port 3, which contains an outgoing antisymmetric Lamb wave. (c) Zoom-in of port 2, which contains a dominant symmetric and low-amplitude antisymmetric Lamb wave.}
    \label{fig:FEMDisplacement}
\end{figure*}

We now utilize both the CMM and FEA of the elastic wave circulator derived in Sec.~\ref{sec:CoupledMode} and Sec.~\ref{sec:FEM}, respectively, to seek circulator and modulation designs that yield a large degree of nonreciprocity at the output port waveguides.
The geometric and material parameters used for both models are shown in Table \ref{table:1}. 
The ring is modeled using nominal material property values of steel, and the material properties of the port waveguides are chosen to be representative of rubber.
This choice in the material property contrast results in a $\bar{Z} = 1\mathrm{e}{-4}$, which induces light damping due to the port waveguides and therefore confines the nonreciprocal behavior to be close to the resonance frequencies of the unmodulated circulator \cite{fleury2015subwavelength}.
A unit input force is prescribed for both models, i.e., $F_0=1$ in Eq.~\eqref{eq:ForcingR}, and the reference displacement is chosen to be $U=8\mathrm{e}{-7}$.

The parameter $\gamma$, which depends on ring geometry, can strongly influence the accuracy of the predicted resonance frequency for the CMM.
The definition of $\gamma$ in Eq.~\eqref{eq:Gamma} is only accurate for long-wavelength modes compared to the circulator thickness, $\lambda/h \ll 1$.
A better approximation to $\gamma$ can be sought by calibrating the CMM with FEA.
This is accomplished by first solving for the eigenvalues of the unmodulated circulator without port waveguides and spatiotemporal modulation, i.e. $(\alpha,\omega_\mathrm{m}, \bar{Z})=(0,0,0)$, using FEA and fitting the parameter $\gamma$ using the equation \cite{graff2012wave}
\begin{equation} \label{eq:GammaFit}
    \gamma = \sqrt{\frac{(\bar{\omega}_n^2-1)(\bar{\omega}_n^2-n^2)-n^2}{n^2(\bar{\omega}_n^2-1) + n^4(\bar{\omega}_n^2-n^2+2)}},
\end{equation}
which is derived by inverting the dispersion relation of the ring for the parameter $\gamma$. 
The parameters in Eq.~\eqref{eq:GammaFit} are the mode number $n$, and $\bar{\omega}_n = \omega_\mathrm{FEA,n}R/c_0$, where $\omega_\mathrm{FEA,n}$ is the resonance frequency for mode $n$ computed with FEA.

In this work, we focus only on the mode $n=4$, which we found to be the mode with the smallest resonance frequency displaying strong nonreciprocity for small modulation parameters.
This results in $\gamma = 0.0149467$, which is used for the remainder of this work.
The normalized Young's modulus of the ring is modulated with the traveling wave form
\begin{equation}
    \bar E(\theta,t) = 1 + \alpha\cos(\theta - \omega_\mathrm{m}t), 
\end{equation}
which results in $\hat{E}^{\pm 1}(\theta) = e^{\pm i \theta}/2$, $\hat{E}^{|p|>1} = 0$ in Eq.~\eqref{eq:YoungsModFourierSeries}.

Figure \ref{fig:FEM_CM_Comparison} shows the drive frequency component ($p=0$ in Eqns.~\eqref{eq:wTotalSoln}-\eqref{eq:vTotalSoln}) of the radial and tangential displacement magnitudes of the ring at the port waveguide locations computed with the CMM and FEA with prescribed modulation parameters $(\alpha, \bar \omega_\mathrm{m}) = (0.2,0.016)$ as a function of the drive frequency. 
We first note that excellent agreement is obtained between the two models.
The degree of nonreciprocity increases as the drive frequency approaches the value $\bar \omega_0 = 0.21604$, where the radial displacement magnitude at port 3 exhibits a null while the radial displacement magnitude at port 2 exhibits a maximum.
The tangential displacement also displays similar behavior, except a maximum in the displacement magnitude is present at port 3 and a minimum in the displacement magnitude is present at port 2. Note, however, that the degree of nonreciprocity for the tangential displacement is not as large as the radial displacement.
The effects of this nonreciprocal ring displacement solution on the waveguide modes will be discussed further below.
The large degree of nonreciprocity shown in Fig.~\ref{fig:FEM_CM_Comparison} is further elucidated by investigating the magnitude of the modal amplitudes from the CMM normalized by the magnitude of the maximum modal amplitude in decibels, which is shown in Fig.~\ref{fig:AmpMatrix}.
We find that the large degree of nonreciprocity at the drive frequency ($p=0$) stems from the interference pattern of the clockwise and counter-clockwise mode ($n=-4,4$), which is qualitatively similar to previous works \cite{fleury2015subwavelength}. 
In addition, harmonic frequencies are generated but are more than $18$ dB lower than the modes generated at the drive frequency. 

\begin{figure}
    \centering
    \includegraphics[width=1\columnwidth]{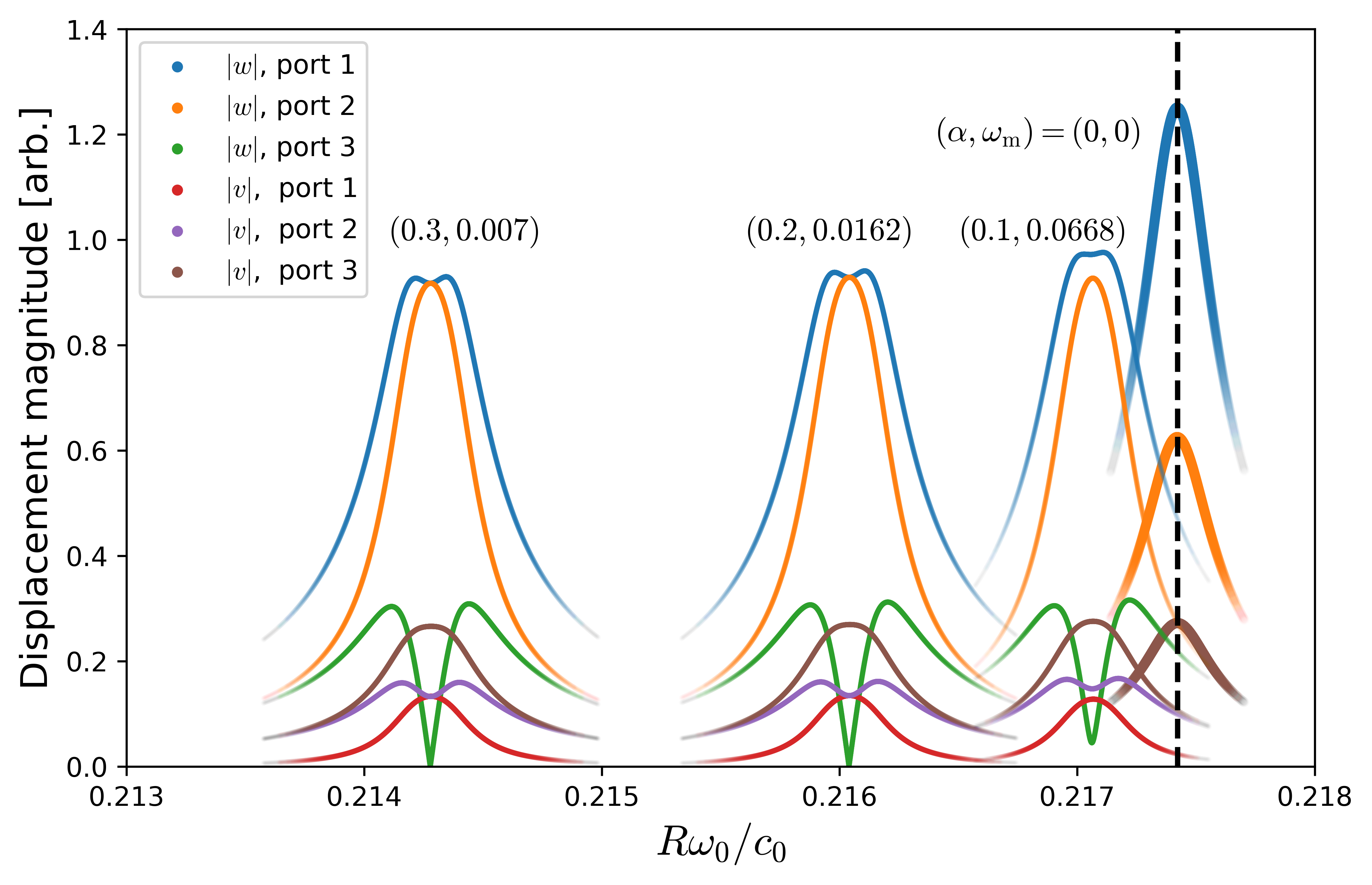}
    \caption{Magnitude of the radial and transverse displacements at the port waveguide locations for various optimal modulation parameters. The modulation parameters are given in parenthesis above each set of curves. The vertical dashed line is the resonance frequency of the unmodulated circulator. The displacements $|w|$ and $|v|$ are identical at ports 2 and 3 for the unmodulated case. The fading of the curves is utilized to minimize the clutter due to overlap of multiple curves on a single plot.}
    \label{fig:OptDisplacement}
\end{figure}
\begin{figure}
    \centering
    \includegraphics[width=1.0\columnwidth]{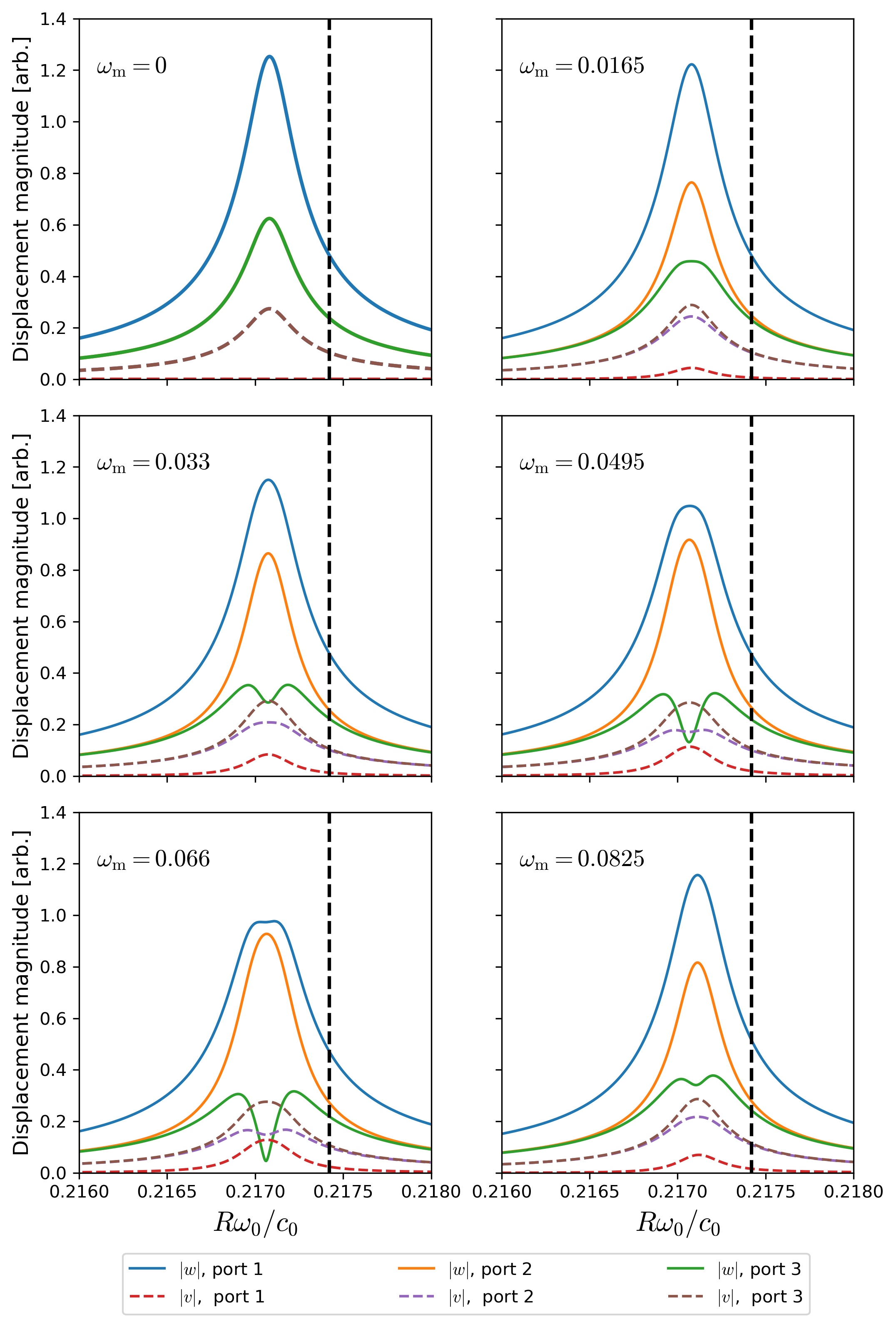}
    \caption{Magnitude of the radial and transverse displacements at the port waveguide locations for various modulation frequencies with fixed $\alpha =0.1$. The vertical dashed line is the resonance frequency of the unmodulated circulator.}
    \label{fig:VaryingOmegaM}
\end{figure}

Figure \ref{fig:FEMDisplacement} shows the displacement solution of the finite element model at the drive frequency $\bar \omega_0 = 0.21604$ for spatiotemporal modulation parameters $(\alpha,\bar \omega_\mathrm{m}) = (0.2, 0.016)$.
Note that transverse displacements of the ring corresponds to an excitation of the antisymmetric ($A_0$) Lamb wave in the waveguide ports, while radial displacements of the ring induce the symmetric ($S_0$) Lamb wave in the waveguide ports.
Therefore, the nonreciprocity of the ring displayed in Fig.~\ref{fig:FEM_CM_Comparison} indicates that the present circulator design acts as a nonreciprocal Lamb wave converter, where the incident symmetric Lamb wave in port 1 is converted to an antisymmetric Lamb wave in port 3 and a dominant symmetric Lamb wave in port 2, as shown in Fig.~\ref{fig:FEMDisplacement}.
While the presence of the null in the radial displacement magnitude at port 3 in Fig.~\ref{fig:FEM_CM_Comparison} indicates efficient nonreciprocal conversion of the incident symmetric Lamb wave to an antisymmetric Lamb wave, the same cannot be said about the tangential displacement magnitude.
As a consequence, port 2 will contain a combination of symmetric and antisymmetric Lamb waves, although the magnitude of the antisymmetric Lamb wave is 17 dB lower than the symmetric Lamb wave for this case.
The further reduction of the antisymmetric Lamb wave in port 2 could potentially be optimized by considering different material properties of the ring and waveguides, as well as three-dimensional effects such as port waveguide vertical thicknesses that are smaller than the ring vertical thickness, which is left for future work. 

We now seek to optimize the nonreciprocal response of the proposed elastic wave circulator by fixing the dimensionless modulation amplitude $\alpha$ and sweeping over the drive and modulation frequencies. The objective function for the minimization problem is the ratio of the absolute square of the radial displacement at port 3 over the radial displacement at port 2, i.e. $|w(\theta_3)/w(\theta_2)|^2$. 
The optimization is carried out using the \textit{minimize} function in the scientific library SciPy for Python \cite{virtanen2020scipy}.
Figure \ref{fig:OptDisplacement} shows the results from the optimization for the modulation depth parameter values of $\alpha = 0.1$, $0.2$, and $0.3$. The unmodulated response is also included as a reference.
We find that as $\alpha$ is increased, smaller values of $\omega_\mathrm{m}$ are needed to obtain a large nonreciprocal response.
In addition, increasing $\alpha$ corresponds to a downward shift in the drive frequency at which the large nonreciprocal response appears.
We also note that the magnitude of the radial displacement at port 2 increases compared to the unmodulated circulator when modulation is active, while the amplitude of the tangential displacement at port 3 stays at nearly the same level.

Finally, the effect of the modulation frequency on the degree of nonreciprocity is further studied by investigating the variation of the ring displacement magnitude as a function of $\omega_\mathrm{m}$ for fixed $\alpha = 0.1$ in Fig.~\ref{fig:VaryingOmegaM}.
We find that the drive frequency at which the large nonreciprocal response appears is not sensitive to changes in the modulation frequency, and is therefore only influenced by $\alpha$.
As the modulation frequency increases the degree of nonreciprocity between the port displacement increases up to a critical value (bottom left panel in Fig.~\ref{fig:VaryingOmegaM}) before the nonreciprocal performance decreases, suggesting that a unique, optimal value of $\omega_\mathrm{m}$ may exist for a given $\alpha$. 

\section{Conclusion}\label{sec:conclusion}
We have developed a semi-analytical CMM and a FEA to study nonreciprocal propagation in an elastic wave circulator in which the Young's modulus is modulated in a traveling wave fashion around a ring-shaped junction with three semi-infinite elastic waveguide ports.
The CMM is based on curved Euler-Bernoulli beam theory and is valid for small curvatures and low frequencies of excitation. 
In contrast, the FEA solves the exact elastodynamic equations and incorporates a mode-matching technique that exactly accounts for outward-propagating Lamb waves in the port waveguides. Both models can be applied to nonreciprocal elastic systems that contain coupling between multiple resonant frequencies as well as multiple wave polarizations.

For the cases considered in this work, we have demonstrated that this system displays a large degree of nonreciprocity near the resonance frequencies of the unmodulated system. In particular, the thin ring acts as a nonreciprocal Lamb wave conversion device, in which an incident symmetric Lamb wave is converted into outgoing symmetric and antisymmetric Lamb waves at the two output port waveguides. The maximum degree of nonreciprocity and the corresponding drive frequency are primarily influenced by the modulation frequency and amplitude, respectively.
We have found excellent agreement between the CMM and FEA for the present case, although we expect the CMM to become less accurate for cases of larger ring and port thicknesses, as well as high frequency modes. Therefore, the FEA is needed to explore the entire parameter space of elastic wave circulator designs.
Further design optimization, as well as finite thickness effects that remove the plane strain assumptions in this work, should be further explored to find circulator designs that yield a large degree of nonreciprocity. In addition, experimental efforts to produce the desired stiffness modulation and measure the nonreciprocal response of the circulator should be conducted to validate the numerical results in this study.


\begin{acknowledgments}
This work supported by National Science Foundation EFRI award No.~1641078 and the Postdoctoral Fellowship Program at Applied Research Laboratories at The University of Texas at Austin.
\end{acknowledgments}

\bibliography{References}
\end{document}